\documentclass[12pt,twoside]{article}
\usepackage{fleqn,espcrc1}
\newcommand{\hedr}{\ensuremath{^{3}\mem{He}}} 
\newcommand{\czw}{\ensuremath{^{12}\mem{C}}} 
\newcommand{\mem}[1]{\ensuremath{\mathrm{ #1}}}   
\newcommand{\nat}[2]{\ensuremath{#1 \cdot 10^{#2}}}     
 \newcommand{\p}{\ensuremath{\mem{p}}}    
\newcommand{\hevi}{\ensuremath{^{4}\mem{He}}} 
\newcommand{\lisi}{\ensuremath{^{7}\mem{Li}}} 
     \newcommand{\bel}{\ensuremath{^{11}\mem{B}}}  
\newcommand{\n}{\ensuremath{\mem{n}}}      
\newcommand{\nvi}{\ensuremath{^{14}\mem{N}}}   
       \newcommand{\emi}{\ensuremath{\mem{e}^\mem{-}}}      
\newcommand{\cdr}{\ensuremath{^{13}\mem{C}}} 
\newcommand{\besi}{\ensuremath{^{7}\mem{Be}}}
\newcommand{\ose}{\ensuremath{^{16}\mem{O}}}    
  \newcommand{\etal}{et~al.\,}           
\newcommand{\kelv}{\ensuremath{\,\rm K}}    
\newcommand{\jahre}{\ensuremath{\, \mathrm{yr}}}  
\newcommand{\apleq}{\ensuremath{\stackrel{<}{_\sim}}}
\newcommand{\apgeq}{\ensuremath{\stackrel{>}{_\sim}}}                 
\newcommand{\abb}[1]{Fig.\,\ref{#1}}          

\usepackage{epsfig}

\newcommand{\AmS}{{\protect\the\textfont2
  A\kern-.1667em\lower.5ex\hbox{M}\kern-.125emS}}

\title{Convective proton and \hedr\ ingestion into helium burning: Nucleosynthesis during a post-AGB thermal pulse}

\author{Falk Herwig\address{University
Potsdam, Potsdam, Germany}$^,$\address{Dept. of Physics and Astronomy, University of Victoria \\ 
Victoria V8W 3P6, B.C., Canada, fherwig@mussel.phys.uvic.ca}$^,$\thanks{This work has been supported by the \emph{Deutsche
            Forschungsgemeinschaft (DFG)} (grant La 587/16). }
        and 
        Norbert Langer\address{Astronomical Institute,  Utrecht University\\ 
Princetonplein 5, NL-3584 CC Utrecht, The Netherlands, N.Langer@astro.uu.nl}}
       
\begin{document}

% typeset front matter
\maketitle

\begin{abstract}
A thermal pulse during the post-AGB phase of stellar evolution may 
lead to a unique mode of light element nucleosynthesis. The stage is set
by the ingestion of the unprocessed envelope material into the
hot He-flash convection zone below.  
%The envelope contains the initial abundance of
%hydrogen and \hedr. = "unprocessed"
If the temperature is sufficiently large and 
the \czw\ abundance high enough (e.g. $T_{\mem{8}}>0.8$, $\mem{X}(\czw) \simeq 
0.4$ and $\mem{X(H)} \simeq \nat{1}{-3}$) protons react faster with \czw\ and form \cdr\ than destroying
\besi. The latter forms by $\alpha$-capture of \hedr\ after an initial
reduction of the \hedr\ abundance to about
$\nat{3}{-5}\mem{X}(\hevi)$ by the ppI reaction
$\hedr(\hedr,2\p)\hevi$ (for $T_{\mem{8}} \simeq 1$).  All \hedr\ is
burned within minutes to weeks depending on the temperature. \besi\ is now present at about the previously
mentioned level of \hedr.
Its further fate  is determined by the 
reactions $\besi(\emi,\nu)\lisi$ and the $\alpha$-capture reactions of 
\besi\ and \lisi. These captures lead to the production of \bel\ which 
in turn is finally destroyed by $\bel(\alpha,\n)\nvi$. The details of
this mechanism of light element production in real stars is
expected to be fairly dependent on the description of mixing.
\end{abstract}

\section{Introduction}
About 20~\% of all post-AGB stars are hydrogen-deficient.
Among them are the Wolf-Rayet type central stars of planetary nebulae 
(PNe) and the PG1159 stars, with their characteristic helium, carbon and
oxygen abundance patterns \cite{koesterke:97b,demarco:97,dreizler:98}. 

 In order to understand the evolutionary
origin and status of hydrogen-deficient objects new stellar evolution
calculations have been recently presented by Herwig
\cite{herwig:00a}. One channel of post-AGB evolution leading to a
hydrogen-deficient surface involves a so-called very late thermal
pulse (VLTP) as first proposed by Fujimoto \cite{fujimoto:77}. During
such an event the entire unprocessed  
envelope is mixed into the region of the ongoing He-flash on the
convective time scale. Corresponding models by Herwig
\etal\cite{herwig:99c} have been constructed using an advanced numerical algorithm
for the treatment of convective nucleosynthesis. It ensures 
consistent abundance profiles and energy generation rates even when
nuclear and
convective time scales are of the same order of magnitude. This is
achieved  by a
simultaneous solution of the nuclear network equations and the
equations of time-dependent convective mixing. 

A group of stars which may be related to the hydrogen-deficient
post-AGB stars are the R\,CrBr stars, of which some are not
only hydrogen-free but at the same time lithium-rich \cite{asplund:00}. The
famous PN central star \emph{Sakurai's object} (V4334 Sgr) resembles
the R\,CrBr stars. However, it showed a rapid evolution of temperature,
luminosity and surface abundances over the last decade
\cite{asplund:99a,duerbeck:00}. Its evolution in the
Hertzsprung-Russell diagram is similar to that of stellar models during
a VLTP.  Its lithium abundance
exceeds the initial solar abundance by 0.5 to 1.0 dex.
This large lithium abundance is surprising given the fragility
of lithium ($T_{\mem{min}}(\lisi + \p) \sim \nat{2}{6}\kelv$).
We have thus investigated the fate of the light elements, in
particular of \hedr, which is present in the envelope at the time of
ingestion into the convectively unstable intershell region of the
ongoing He-flash beneath. Here, we present a new reaction channel which is
to be considered as\emph{ hot H-deficient \hedr\ burning.}
  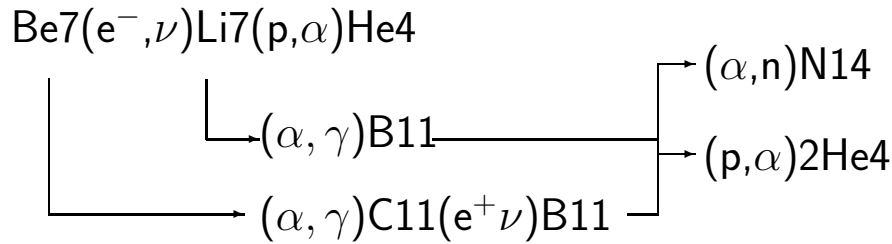
\begin{figure}[t]
      \hspace{2cm}
      \begin{minipage}[h]{4cm}
      \setlength{\unitlength}{4144sp}%
\begingroup\makeatletter\ifx\SetFigFont\undefined%
\gdef\SetFigFont#1#2#3#4#5{%
  \reset@font\fontsize{#1}{#2pt}%
  \fontfamily{#3}\fontseries{#4}\fontshape{#5}%
  \selectfont}%
\fi\endgroup%
\begin{picture}(4140,1389)(1306,-1708)
%\thinlines
\put(2476,-736){\line( 0,-1){360}}
\put(2476,-1096){\vector( 1, 0){315}}
\put(1531,-736){\line( 0,-1){810}}
\put(1531,-1546){\vector( 1, 0){1170}}
\put(3826,-1096){\line( 1, 0){1350}}
\put(5176,-1096){\line( 0, 1){450}}
\put(5176,-646){\vector( 1, 0){225}}
\put(4996,-1546){\line( 1, 0){180}}
\put(5176,-1546){\line( 0, 1){450}}
\put(5176,-1096){\line( 0,-1){180}}
\put(5176,-1276){\line( 0, 1){180}}
\put(5176,-1096){\line( 0,-1){ 90}}
\put(5176,-1186){\vector( 1, 0){225}}
\put(1306,-511){\makebox(0,0)[lb]{\smash{\SetFigFont{17}{20.4}{\sfdefault}{\mddefault}{\updefault}Be7(e$^-$,$\nu$)Li7(p,$\alpha$)He4}}}
\put(2791,-1141){\makebox(0,0)[lb]{\smash{\SetFigFont{17}{20.4}{\sfdefault}{\mddefault}{\updefault}($\alpha,\gamma$)B11}}}
\put(2791,-1636){\makebox(0,0)[lb]{\smash{\SetFigFont{17}{20.4}{\sfdefault}{\mddefault}{\updefault}($\alpha,\gamma$)C11{(e$^+\nu$)}B11}}}
\put(5446,-736){\makebox(0,0)[lb]{\smash{\SetFigFont{17}{20.4}{\sfdefault}{\mddefault}{\updefault}($\alpha$,n)N14}}}
\put(5446,-1276){\makebox(0,0)[lb]{\smash{\SetFigFont{17}{20.4}{\sfdefault}{\mddefault}{\updefault}{(p,$\alpha$)}2He4}}}
\end{picture}
      \end{minipage}
    \caption{\label{fig:alpha-cap} The ppII-chain and the additional reactions 
      considered for ingestion of light elements into He-burning
      conditions. At relevant temperatures around $T_\mem{8} \sim 1.5$
      the $\alpha$-capture on \lisi\ and \bel\ are four to five and
      three orders of magnitude larger than the $\cdr\ + \alpha$
      reaction, respectively. \besi\ is more similar to the carbon
      isotopes with its $\alpha$-capture being about tenfold that of
      \cdr\ and the \p-capture exceeding that of \czw\ by a factor of
      five. The \emi-capture of \besi\ has a time scale of $0.1 \jahre$.}
  \end{figure}

\section{Nucleosynthesis of light elements during a VLTP}
Any lithium production in a stellar environment is the result of the
ppII reaction chain which is based on a readily available \hedr\
reservoir. In the event of ingestion of unprocessed envelope material
into the active and convectively unstable helium shell, protons are
mainly captured by \czw\ and no production of \hedr\ is
possible. However, the \hedr\ which is preserved in the envelope will be
processed by the well known ppI-chain  or by capturing an
$\alpha$-particle. Any  
\besi\ produced this way is obviously threatened by proton captures
(ppIII-chain) as well as \emi-captures (ppII-chain). In addition, due
to the high   
temperature in the helium shell ($T_8 \simeq 1 \dots 1.5 ...2.5$)
$\alpha$-captures on any light isotopes had to be taken into account as
well (\abb{fig:alpha-cap}). The reaction rates used in the present study
have been taken from
the \texttt{NACRE} compilation \cite{angulo:99} except the
$\bel(\alpha,n)\nvi$ rate which was only available from Caughlan \&
Fowler \cite{caughlan:88}.

  \begin{figure}[t]
    \begin{center}
      \epsfxsize=\textwidth
      \epsfbox{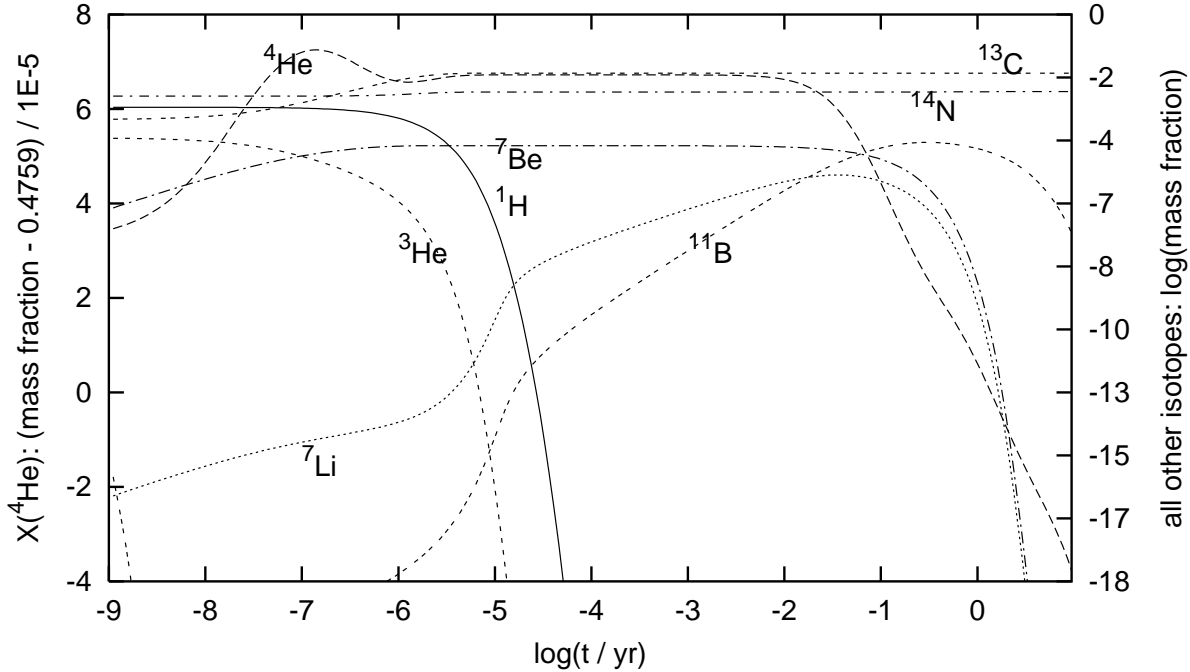}
    \end{center}
    \caption{\label{fig:one-zone}  Time evolution of isotopic abundances
      for constant temperature $T_8=1.5$ and density $\rho =
  1000 \mem{g}/\mem{cm^3}$. The initial abundances resemble the
  AGB envelope composition ($\log X(^3\mem{He}) = -3.9$) with modifications: \hevi, \czw\ and \ose\
  have the intershell abundance of AGB models with overshoot
  (0.47/0.40/0.11) and the initial hydrogen abundance has been chosen
  as $\log X(\mem{H}) = -3$ to account for the dilution effect of the
  mixing (all abundances are given as mass fractions).}
  \end{figure}
In order to demonstrate the principal aspects of the hot H-deficient
\hedr\ burning we have constructed a series of one-zone nuclear
network models which include the CNO isotopes, the pp-chains with the
additions shown in \abb{fig:alpha-cap} and helium burning, for typical
initial conditions. A representative example is shown
in \abb{fig:one-zone}.

During the first second of the evolution, the ppI reaction
$\hedr(\hedr,2\p)\hevi$ prevails until
$X_{\hedr} < \frac{3}{4}\frac{\sigma(\hevi + \hedr)}{\sigma(\hedr + \hedr)}
X_{\hevi} \sim \nat{1.2}{-5} $. Then the ppII reaction
$\hedr(\alpha,\gamma)\besi$ is dominating the nuclear burning of
\hedr. \besi\ is built up to a mass fraction of $\sim 10^{-5}$ which
corresponds to the \hedr\ abundance at which the \besi\ production
takes over. The half-life of protons against destruction by capture of
\besi\ is $2.2\mem{d}$, much longer than the rapid capture of protons
by the abundant \czw\ with a time scale of $35 \mem{sec}$. Therefore
\besi\ produced under these conditions will not be destroyed by proton
captures because the protons are quickly locked up in the \cdr\
abundance. \besi\ will  be transformed into \lisi\ which peaks at
$\log t/[yr] = -1$. Then, $\alpha$-captures are processing \lisi\ into
\bel. If the temperature is significantly larger ($T_8 \apgeq 2$) 
the time scale of $\alpha$-capture on \lisi\ is smaller than that of
\emi-capture by \besi ~ ($\tau_{\alpha}(\lisi) <
\tau_{e^-}(\besi)$). In that case no significant \lisi\ abundance can
build up because any \emi-capture by \besi\ is almost immediately
followed by an $\alpha$-capture. If, on the other hand,  $T_\mem{8} \apleq
0.8$ then  $\tau_{\besi}(\p) < \tau_{\czw}(\p)$ and protons are not
efficiently captured by \czw\ but instead destroy \besi\ (the
common ppIII-chain). Thus, the proposed mechanism of hot H-deficient
\hedr\ burning can only work in a temperature window from
$T_\mem{8}=0.8$ to $T_\mem{8}=2$.

In a second step we have constructed models of coupled convective
mixing and nuclear burning, using the same numerical method as
described in Herwig \cite{herwig:00a}. Following a post-processing
approach the chemical evolution has been simulated. The hydrostatic
input data have been taken from the VLTP model sequence of Herwig
\etal\cite{herwig:99c}. It turns out that \lisi\ is produced in the
upper part of the He-flash convection zone, after the sudden energy
release of hydrogen-burning leads to the well known split of this
convective region. Without that short radiative separation of the flash
convection zone the nuclear processes are dominated by the hottest
temperatures in the convective zone ($T_\mem{8}=2.5$) which exceed the
limit at which lithium production is possible as described above. At
the end of the post-processing simulation the
lithium abundance in the region immediately below the surface exceeds
a mass fraction of $10^\mem{-8}$ which would be in accordance with the
observed abundance of \emph{Sakurai's} object. 

\section{Conclusions}
We have presented a new mode of light element nucleosynthesis which
likely occurs when unprocessed material is ingested into an active and
convectively unstable helium-shell, e.g.\ during a very late post-AGB
thermal pulse. We have, however, not yet modelled in detail the mixing
processes which ultimately bring the lithium into the
atmosphere. Given the fragile constitution of lithium, most other
sources of lithium in stars like \emph{Sakurai's} object can
be ruled out. We thus conclude that the proposed mechanism might be
responsible for some of the lithium abundance anomalies observed in
stars.

%\bibliography{astro}

\begin{thebibliography}{11}
\expandafter\ifx\csname natexlab\endcsname\relax\def\natexlab#1{#1}\fi
\expandafter\ifx\csname url\endcsname\relax
  \def\url#1{{\tt #1}}\fi

\bibitem{koesterke:97b}
L.~{Koesterke} and W.~R. {Hamann}.
\newblock A\&A 320 (1997) 91.

\bibitem{demarco:97}
O. {De Marco}, P.~J. {Storey}, and M.~J. {Barlow}.
\newblock MNRAS 297 (1998) 999.

\bibitem{dreizler:98}
S.~Dreizler and U.~Heber.
\newblock A\&A 334 (1998) 618.

\bibitem{herwig:00a}
F.~Herwig.
\newblock In T.~Bl\"ocker, R.~Waters, and B.~Zijlstra, editors,  Low mass
  Wolf-Rayet Stars: origin and evolution, Ap\&SS (2000)
\newblock in press, astro-ph/9912353.

\bibitem{fujimoto:77}
M.~Y. Fujimoto.
\newblock PASJ 29 (1977) 331.

\bibitem{herwig:99c}
F.~Herwig, T.~Bl\"ocker, N.~Langer, and T.~Driebe.
\newblock A\&A 349 (1999) L5.

\bibitem{asplund:00}
M.~{Asplund}, B.~{Gustafsson}, D.~L. {Lambert}, and N.~K. {Rao}.
\newblock A\&A, 353 (2000) 287.

\bibitem{asplund:99a}
M.~{Asplund}, D.~L. {Lambert}, T.~{Kipper}, D.~{Pollacco}, and M.~D.
  {Shetrone}.
\newblock A\&A 343 (1999) 507.

\bibitem{duerbeck:00}
H.~W. Duerbeck, W.~Liller, and S.~Benetti etal.
\newblock AJ, 119 (2000) 2360.

\bibitem{angulo:99}
C.~{Angulo}, M.~{Arnould}, and {Rayet, M. et al.}
\newblock Nucl.\ Phys.\ A 656 (1999) 3.

\bibitem{caughlan:88}
G.~R. Caughlan and W.~A. Fowler.
\newblock Atom.\ Data Nucl.\ Data Tables 40 (1988) 283.



\end{thebibliography}

\end{document}